\documentstyle[12pt]{article}

\begin{document}

\begin{titlepage}
\title{Spectrum of an Elliptic Free Fermionic Corner
Transfer Matrix Hamiltonian}

\author{R. Cuerno \\
{\it Instituto de Matem\'aticas y F\'{\i}sica Fundamental, CSIC} \\
{\it Serrano 123, E--28006 Madrid, SPAIN} \\
{\tt e-mail:IMTRC59@CC.CSIC.ES}}
\date{}

\maketitle

\begin{abstract}
The eigenvalues of the Corner Transfer Matrix Hamiltonian
associated to the elliptic $R$ matrix of the eight vertex free
fermion model are computed in the anisotropic case for magnetic
field smaller than the critical value.
An argument based on generating functions is given, and
the results are checked numerically. The spectrum consists of
equally spaced levels.
\end{abstract}

\vskip-13.0cm
\rightline{{\bf IMAFF 93/13}}
\rightline{{\bf August 1993}}
\vskip2cm

\end{titlepage}

\indent
In recent times a great attention has been increasingly
paid to the computation of correlation functions for integrable
models in two dimensions. Following the approach of the Kyoto
school it has been possible not only to diagonalize hamiltonians
(as the XXZ model \cite{DFJMN} and its higher spin analogs \cite{IIJMNT}
in the thermodynamic limit for the corresponding massive regimes) by
means of the representation theory of
quantum affine algebras, but also to extend the
arguments to cases for which no quantum group like structure has
been identified yet, namely the zero field eight--vertex (8V) model \cite{JMN}.
A prominent role is played in this approach by the Corner Transfer Matrix
Hamiltonian (CTMH), which in the quantum group invariant cases acts as
a derivation in the quantum affine algebra (its physical
interpretation is to be the lattice analog of the generator of
Lorentz boosts \cite{TT}). A salient feature of this operator is that its
spectrum consists of equally spaced levels \cite{Bax}. Remarkably
there exists an 8V model (the elliptic free
fermionic solution of the eight vertex Yang--Baxter equation \cite{F,SUAW,BS})
for which a sort of affine quantum group structure $\widehat{CH}_q(2)$ has
been identified, in which the $R$ matrix acts as an intertwiner
\cite{CGLS}. The associated row--to--row hamiltonian
is an (anisotropic) XY model in a magnetic field. This letter is a
first step in the program of employing quantum group invariance
to compute correlation functions in this 8V--like model.
Namely it is shown here that the spectrum of the corresponding CTMH in the
anisotropic regime\footnote{for values of the magnetic field
smaller than the critical one. See below.} is given essentially by integer
numbers, which makes it reasonable to try for it an interpretation
as a derivation--like operator in $\widehat{CH}_q(2)$.
This CTMH has already been studied in the context of the so called
master symmetries \cite{A} and its spectrum has
been studied through the use of orthogonal polynomials \cite{TP1,ET}. However
in the latter references only the trigonometric (isotropic) and hyperbollic
(disorder line) limits have been worked out, since the more
general case leads to 5--term recurrence relations whose
link to the 3--term recurrences characteristic of
orthogonal polynomials \cite{Chi} is unclear. In the present
letter the use of both a convenient parametrization of the model and
an argument following Davies \cite{Davies},
dealing with generating functions enables us to identify the correct form
of the eigenvalues of the CTMH also in a more general case.
We compare the results to some numerical computations and then end
with some comments and suggestions.

The most general solution for the elliptic free fermionic 8V
$R$ matrix is, in the parametrization of ref. \cite{BS}:
\begin{eqnarray}
R_{00}^{00}(u)&=&1-e(u)e(\psi_1)e(\psi_2) \;\;\;\;\; R_{11}^{11}(u)
= e(u)-e(\psi_1)e(\psi_2) \nonumber \\
R_{01}^{01}(u)&=&e(\psi_2)-e(u)e(\psi_1) \;\;\;\;\;\; R_{10}^{10}(u) =
e(\psi_1)-e(u)e(\psi_2) \label{4.1} \\
R_{01}^{10}(u)&=&R_{10}^{01}(u)=(e(\psi_1){\rm sn}(\psi_1))^{1/2}
(e(\psi_2){\rm sn}(\psi_2))^{1/2}(1-e(u))/{\rm sn}(u/2) \nonumber \\
R_{00}^{11}(u)&=&R_{11}^{00}(u)=-{\rm i}k(e(\psi_1){\rm sn}(\psi_1))^{1/2}
(e(\psi_2){\rm sn}(\psi_2))^{1/2}(1+e(u)){\rm sn}(u/2) \nonumber
\end{eqnarray}

\noindent
with $e(u)$ the elliptic exponential:
\begin{equation}
e(u)={\rm cn}(u)+{\rm i}\;{\rm sn}(u) \nonumber
\end{equation}

\noindent
$k$ the elliptic modulus and ${\rm cn}(u)$, ${\rm sn}(u)$ the Jacobi
elliptic functions of modulus $k$. In the sequel we will set
\begin{equation}
\psi_1=\psi_2 \equiv \psi \nonumber
\end{equation}

\noindent
since this is the interesting case from the hamiltonian point of
view. Up to a normalization constant the row--to--row
hamiltonian acting on the $j$--th and $j+1$--th spins is
\begin{equation}
H_{j,j+1} = {\rm i} \left. \frac{d R_{j,j+1}(u)}{du}\right|_{u=0}
= (1+ \Gamma) \sigma_j^x \sigma_{j+1}^x +
(1- \Gamma) \sigma_j^y \sigma_{j+1}^y + h(\sigma_j^z +
\sigma_{j+1}^z)
\nonumber
\end{equation}

\noindent
The $\sigma$'s are the spin 1/2 Pauli matrices and
\begin{equation}
\Gamma \equiv k \; {\rm sn}(\psi) \;\; , \;\;
h \equiv {\rm cn}(\psi) \label{def}
\end{equation}

\noindent
For real $\psi$ and $0 < k \leq 1$, we stay in the Lieb--Schultz--Mattis
(LSM) unit square
\cite{LSM,BM}
\begin{equation}
0 < \Gamma \leq 1 \;\;,\;\; 0\leq h <1 \label{box}
\end{equation}

\noindent
We will take $h$ and $k$ as our physical independent parameters,
(see Figure 1) though some expressions will be left in terms
of $\psi$. A crucial point is the elliptic function dependence
on this parameter shown in (\ref{def}).

To define the CTM first construct a finite row of vertices by
(see for instance \cite{IT})
\begin{equation}
G_j^{(n)}(u) \equiv R_n(u) R_{n-1}(u) \cdots R_j(u)
\nonumber
\end{equation}

\noindent
where $R_j$ is the Boltzmann weight matrix correponding to the
$j$--th vertex in the row. Then construct a wedge shaped region
by piling up rows of vertices
\begin{equation}
A_N(u) = G_1^{(N)}(u) G_2^{(N)}(u) \cdots G_N^{(N)}(u) \label{CTM}
\end{equation}

\noindent
We shall consider the thermodynamic limit $N \rightarrow
\infty$ for which Baxter has shown \cite{B} that the normalized (by its
highest eigenvalue) CTM $A_N(u)$ is a one parameter commuting family
written as
\begin{equation}
A_n(u) = \exp (- u H_{CTM}) \nonumber
\end{equation}

\noindent
Expanding (\ref{CTM}) for small $u$ one finds for the CTMH
\begin{equation}
H_{CTM} = \sum_{j=1}^{\infty} j H_{j,j+1} \label{CTMH}
\end{equation}

\noindent
This is the operator which we intend to diagonalize.
Since the model is free fermionic we find it convenient to
search for eigenvectors of (\ref{CTMH}) which are linear combinations of
the Jordan--Wigner fermions \cite{HR}
\begin{eqnarray}
& \tau_j^{x,y} \equiv \frac{1}{\sqrt{2}} e^{\frac{\pi {\rm i}}{2}
\sum_{k=1}^{j-1} (\sigma_k^z- 1)} \sigma_j^{x,y} & \nonumber \\
& & \label{tau} \\
& \left\{ \tau^a_j, \tau^b_k \right\} = \delta_{jk} \delta_{ab}
& \nonumber
\end{eqnarray}

\noindent
In terms of these
\begin{equation}
H_{CTM} = - 2 {\rm i} \sum_{j=1}^{\infty} j \left\{ (1+\Gamma)
\tau_{j}^{y} \tau_{j+1}^{x} - (1-\Gamma) \tau_{j}^{x}
\tau_{j+1}^{y} + h (\tau_{j}^{x} \tau_{j}^{y} + \tau_{j+1}^{x}
\tau_{j+1}^{y}) \right\} \nonumber
\end{equation}

\noindent
So look for states
\begin{equation}
\psi(l) \equiv \sum_{j=1}^{\infty} \left\{ A^-_{l,j} \tau^x_j +
A^+_{l,j} \tau^y_j \right\} \label{psil}
\end{equation}

\noindent
such that
\begin{equation}
\left[ H_{CTM}, \psi(l) \right] = \lambda_l \psi(l) \nonumber
\end{equation}

\noindent
$l$ is some index which labels eigenstates. We get 2 sets of
equations for the numerical coefficients in (\ref{psil})
($\tilde{\lambda}_l \equiv -\lambda_l/2$, $\gamma_{\pm} \equiv 1 \pm
\Gamma$):
\begin{equation}
\gamma_{\pm} (m-1) A^{\pm}_{l,m-1} - h (2m-1) A^{\pm}_{l,m} + \gamma_{\mp} m
A^{\pm}_{l,m+1} = \pm {\rm i} \tilde{\lambda}_l A^{\mp}_{l,m}
\label{ABdisca}
\end{equation}

\noindent
Note that if we substitute any of the (\ref{ABdisca})
into the other we get a 5--term recurrence for either $\{ A_{l,j}^{\pm} \}$.
Such relation readily becomes 3--term
like if we set $\Gamma^2=1$ (Ising model \cite{Davies,TP2}) or $h=0$
(doubled Ising model \cite{IT,TP2}). Also the trigonometric ($k=0$)
and the hyperbollic ($k=1$, disorder line $\Gamma^2 + h^2=1$)
limits of this model correspond in essence to three term recurrences which
can be solved by means of orthogonal polynomials \cite{TP1}. We
want to solve for the anisotropic case (\ref{box}), so define the functions
\begin{equation}
A^{\pm}_l(t) \equiv \sum_{j=1}^{\infty} t^j \; A^{\pm}_{l,j}
\label{gen}
\end{equation}

\noindent
We get the differential equations
\begin{equation}
(\gamma_{\pm} t^2 - 2 h t + \gamma_{\mp}) \frac{d A^{\pm}_l(t)}{dt}
+ (h- \gamma_{\mp} t^{-1}) A^{\pm}_l(t) = \pm {\rm i} \tilde{\lambda}_l
A^{\mp}_l(t)
\nonumber
\end{equation}

\noindent
Introducing the integrating factors
\begin{equation}
f_{\pm}(t) = \frac{t}{(\gamma_{\pm} t^2 -2ht + \gamma_{\mp})^{1/2}}
\nonumber
\end{equation}

\noindent
such that
\begin{equation}
A^{\pm}_l(t) = \alpha^{\pm}_l(t) f_{\pm}(t)
\nonumber
\end{equation}

\noindent
we are left with the system
\begin{equation}
(\gamma_{+} t^2 -2ht + \gamma_{-})^{1/2} (\gamma_{-} t^2 -2ht +
\gamma_{+})^{1/2} \; \frac{d\alpha_l^{\pm}}{dt} =
\pm {\rm i} \tilde{\lambda}_l \alpha^{\mp}_l(t)
\label{sist}
\end{equation}

\noindent
Changing variables to
\begin{equation}
s = \int \frac{dt}{(\gamma_{+} t^2 -2ht + \gamma_{-})^{1/2} (\gamma_{-}
t^2 -2ht + \gamma_{+})^{1/2}} \label{inte}
\end{equation}

\noindent
the most general solution to (\ref{sist}) becomes
\begin{eqnarray}
\alpha^+_l(s) & = & a^{(l)}_1 e^{s\tilde{\lambda}_l}
+ a^{(l)}_2 e^{-s\tilde{\lambda}_l} \nonumber \\
\alpha^-_l(s) & = & - {\rm i} (a^{(l)}_1 e^{s\tilde{\lambda}_l}
- a^{(l)}_2 e^{-s\tilde{\lambda}_l}) \label{sol}
\end{eqnarray}

\noindent
where $a^{(l)}_{1,2}$ are arbitrary integration constants.
Due to the freedom in an overall constant we will be interested
in the value of their ratio $\gamma^{(l)} = a^{(l)}_2/a^{(l)}_1$.
So the whole issue resides in solving for the integral
(\ref{inte}). Notice that in the generic case the integrand has
four different poles (they appear as complex conjugate pairs) at
\begin{equation}
\frac{{\rm cn}(\psi) \pm {\rm i} k'{\rm sn}(\psi)}{1
+ k {\rm sn}(\psi)} \;\; , \;\; \frac{{\rm cn}(\psi)
\pm {\rm i} k'{\rm sn}(\psi)}{1- k {\rm sn}(\psi)}
\end{equation}

\noindent
where $k'$ is the conjugate elliptic modulus $k^2+k'^2=1$.
We recall the reader that these values coincide precisely with
those of the central elements for quantum Clifford--Hopf
$CH_q(2)$ invariant open chain Hamiltonians \cite{CGR}.

An adequate change of variables is \cite{G}
\begin{equation}
y = \frac{\gamma_{+} t^2 -2ht + \gamma_{-}}{\gamma_{-}
t^2 -2ht + \gamma_{+}}
\label{cambio}
\end{equation}

\noindent
Notice that it breaks down in the trigonometric limit, which
should be treated separately in this approach. Being $y_{\pm} =
\frac{1\pm k}{1\mp k}$ the maximum and the minimum value of $y$
respectively we get
\begin{equation}
s = \frac{1}{2 k'{\rm sn}(\psi)} \int_y^{y_+} \frac{dy}{(y
(y_+-y) (y-y_-))^{1/2}} \nonumber
\end{equation}

\noindent
and then \cite{Byrd}
\begin{equation}
s = \frac{1}{(1+k) |{\rm sn}(\psi)|} {\rm tn}^{-1} \left[
\left(\frac{1+k}{1-k} \right)^{1/2} \frac{t\; r_- - r_+}{t\; r_+ - r_-},
\frac{2 k^{1/2}}{1+k} \right] \nonumber
\end{equation}

\noindent
with $r_{\pm} = (1 \pm {\rm sn}(\psi))^{1/2}$ and ${\rm
tn}^{-1}(u,k)$ an inverse Jacobian elliptic function
of modulus $k$. Now we want
that the functions (\ref{gen}) generate normalizable
states when $N \rightarrow \infty$, so we require that they are free of
singularities inside the unit circle $|t|=1$ \cite{Davies}. This
completely fixes the values of $\tilde{\lambda}_l$ and
$\gamma^{(l)}$, with the result
\begin{equation}
\lambda_l = \frac{2 \pi l}{K'(k)} |{\rm sn}(\psi)| \;\;\;\;\;\; l=0,1,2,\ldots
\label{res}
\end{equation}

\noindent
where $K'(k)$ is the complete elliptic integral of the first kind
of modulus $k'$. This spectrum reduces
to its known values in the zero field ($h=0$, $\psi=K(k)$) \cite{IT}
and disorder line \cite{TP1} cases.

Finally we present some numerical confirmation of the above
result for the spectrum. Since our CTMH is a quadratic form in
fermionic operators we can try to diagonalize it by finding the
eigenvalues $\lambda_l^2$ of the $N \times N$ LSM matrix \cite{LSM}
\begin{equation}
(A-B)(A+B)= M M^t \label{mat}
\end{equation}

\noindent
where $M$ is the following tridiagonal matrix
\begin{equation}
M = \left( \begin{array}{ccccccc}
h (0+1) & \gamma_- & 0 & \cdots & 0 & 0 & 0 \\
\gamma_+ & h (1+2) & 2 \gamma_- & \cdots & 0 & 0 & 0 \\
0 & 2 \gamma_+ & h (2+3) & \cdots & 0 & 0 & 0 \\
\vdots & \vdots & \vdots & \ddots & \vdots & \vdots &
\vdots \\
0 & 0 & 0 & \cdots & (N-2) \gamma_+ & h (N-1 + N-2) & (N-1) \gamma_- \\
0 & 0 & 0 & \cdots & 0 & (N-1) \gamma_+ & h (N-1) \\
\end{array}
\right)
\nonumber
\end{equation}

\noindent
(\ref{mat}) is a pentadiagonal matrix as mentioned in
\cite{TP1}, and the eigenvalue equation for it leads to the
same 5--term recurrence as (\ref{ABdisca}).
We have diagonalized it numerically for
dimensions up to $N=500$, and found perfect agreement of the lowest
eigenvalues with the formula (\ref{res}) for any values of $k$,
$h$ in the unit square (\ref{box}). In
Fig. 2 we have plotted the lowest eigenvalues for fixed
magnetic field $h$ and let the anisotropy $\Gamma$ vary with
the elliptic modulus $k$; we find good fit of the theoretical
coefficient $C(\psi,k) \equiv 2 \pi |{\rm sn}(\psi)|/K'(k)$.
This is also shown in Table 1.
\begin{table}
\begin{center}
\begin{tabular}{|c|c|c|c|}
\hline
k & h & $C(\psi,k)_{theor}$ & $C(\psi,k)_{num}$ \\
\hline \hline
$10^{-2}$ & 0.5 & 0.908 & 0.91 \\
0.1 & 0.5 & 1.47238 & 1.47238 \\
1 & 0.5 & 3.46410 & 3.46410 \\
\hline
0.5 & 0.1 & 2.89898 & 2.89898 \\
0.5 & 0.5 & 2.52324 & 2.52324 \\
0.5 & 0.999 & 0.1302 & 0.1304 \\
\hline
\end{tabular}
\caption{Numerical vs. theoretical values of $C(\psi,k)$.}
\end{center}
\end{table}
We have found poorer convergence in the cases of small
$\psi$ or $k$ (for fixed value of the other parameter), since
they both correspond to approaching the isotropic limit $\Gamma =0$
(see Fig. 1) which is critical in this parametrization \cite{TP1}.

In conclusion we have found that the CTMH for the elliptic free
fermionic eight vertex model has a spectrum given by integer
values also in an anisotropic range of its parameters. This leads
us to think of its relation to some
derivation--like operator for $\widehat{CH}_q(2)$ (see \cite{A}
for its effective role as a derivation in the context of master
symmetries), and makes it interesting to consider the problem of
the computation of correlation functions in this 8V model
using quantum group representation theory. This will be the subject
of further study. It would be also interesting to complete the
proof of orthogonality and completeness of the basis of
eigenstates that has been generated here \cite{Davies}, which would involve
the use of identities among elliptic functions,
and to think of the pole structure of the
quasimomentum--rapidity like transformation (\ref{inte}) in
connection with quantum group invariance. Other interesting formal aspect
is the relation (if any) of the 5--term recurrence with the 3--term
recurrences satisfied by orthogonal polynomials. The theory of $q$--deformed
orthogonal polynomials (see for instance \cite{Koor}) might be useful in
this. Finally since the trigonometric limit of the $R$
matrix (\ref{4.1}) is related to $N=2$ SUSY (see \cite{Esp} for
the more general $\widehat{CH}_q(D)$ case, $D$ even) we are led to
speculate whether the study of the algebraic properties of the
CTMH we are considering could lead to the definition of a
lattice analog of an $N=2$ superconformal algebra, much in the
same way as \cite{IT} constructed a lattice analog of Virasoro
($N=0$) for the free fermionic point of Baxter's solution of the
zero field 8V model \cite{B} (our $h=0$ case).

\vspace{1cm}
{\bf Acknowledgments}

The author is pleased to thank A. Berkovich,
and G. Sierra for many discussions, suggestions, and for
encouragement, A. Gonz\'alez--Ruiz for comments and V. R. Velasco
for access to the computing facilities of his group. This research
has been supported by the Spanish Ministry of Education and Science
through predoctoral fellowship PN89--11798388.

\newpage
\subsection*{Figure Captions.}

\vspace{.5mm}
\begin{itemize}

\item[Fig. 1.] $(h, k)$ parameter space. Our computation applies
away from the two isotropic border lines $h=1$, $k=0$.

\item[Fig. 2.] Lowest eigenvalues of $M M^t$ for fixed $h=0.5$.
Lines are guides for the eye.

\end{itemize}

\end{document}